# The physical basis of the zero charge flux constraint in gyro kinetics as local momentum conservation, by analogy to the classical Coulomb collisional case

M. Kotschenreuther
University of Texas at Austin

## I. Introduction

Constraints arising from conservation laws play a crucial role in many physical processes. Constraints are often the underlying reason why free energy cannot relax rapidly. A particularly important constraint for magnetically confined plasma is the condition of zero charge flux across field lines. This constraint has long been recognized as arising for the case of Coulomb collisional transport processes, such as classical transport in many geometries, and neoclassical transport in geometries with a symmetry. The same constraint also arises in an extremely different transport process: gyro-kinetic instabilities. For the instabilities, this constraint has recently been found to be a crucial reason why instabilities are weak enough to allow Transport Barriers to be sustained. And TBs are an amazing physical phenomenon in non-equilibrium thermodynamics that should be understood: over a few cm, huge sustained temperature differences are observed in magnetically confined plasmas, that exceeding the temperature difference from the core to the edge of the sun over ~ $10^6$ km. The fact that the same constraint arises for the apparently totally unrelated cases of classical transport and gyro-kinetic transport, is a clue that it may have a general underlying basis. Our purpose here is to understand this.

The zero charge flux constraint is

$$0 = \sum_s q_s \vec{\Gamma}_{\perp s} \qquad \text{Eq(1)}$$

Eq(1) is already well-known in certain, apparently very different limits and cases. Our aim is to show that the basis of these is in fact the same, and, the condition eq(3) applies far more generally. Two of the well-known limits are

1) Classical transport from Coulomb collisions. These have a scale of order the Debye length.

2) Gyrokinetic fluctuation transport. For a limiting case of this transport, eq(1) is well known to hold in the following limiting case:

   A. Electrostatic fluctuations, so that the only motions across field lines are ExB drifts. (Note that this is a significant restriction since, in realistic turbulence, magnetic transport can be important, e.g., from stochastic magnetic field lines.)
   B. Small gyro-radius, so that the gyro-orbit average of the ExB drift is identical for all species. (Note that this is also a strong restriction since, in general in the gyrokinetic equation, when $k_\perp \rho_i$ is significant, the gyro-orbit average ExB drift is significantly different for the ions and electrons. )

C. Vanishing Debye length, implying quasi-neutrality of the fluctuations: $0 = \sum q_s \delta n_s$. (This is not be true for some important gyrokinetic fluctuations, such as ETG modes. These are at the scale of the electron gyro-radius, which is often the same order as the Debye length.)

D. When all the limiting conditions of A-C apply, the perpendicular flux is $\vec{\Gamma}_{\perp s} = (\vec{E} \times \vec{B}/B^2) \delta n_s$ ; then eq(3) is satisfied at every point due to quasi-neutrality, $\sum q_s \delta n_s = 0$.

These apparently *extremely* different cases 1 and 2 are, in fact, just specific instances of the same fundamental physical dynamic. And because of this general dynamic, gyrokinetic fluctuation transport satisfies eq(1) without any of the restricting conditions A-C. The physical basis for eq(1) has been well understood for classical transport, and this is an extremely valuable and clarifying guide for the gyrokinetic case.

As mentioned above, constraints often imply that free energy cannot relax rapidly. This applies in both the gyrokinetic case and the Coulomb collisional transport case, with surprising similarities and important differences. For classical transport, it is well known that "momentum conservation" implies that density gradients relax much more slowly than temperature gradients. This is because self-interactions of one species (ions in this case) can be very strong, giving high energy transport. However, momentum exchange between electrons and ions is necessary to allow a density flux, according to eq(1), and this is lower by $\sim (m_e/m_i)^{\frac{1}{2}}$. For Coulomb collisions, the source of the "fluctuations" is the unavoidable fact that plasma is composed of discrete particles. Gyrokinetic fluctuations arise from instabilities. These instabilities depend upon the ability of collective modes to access equilibrium free energy to grow. As shown in detail elsewhere, eq(1) can make it impossible for strong instabilities to arise, even in the presence of temperature gradients, when there are sufficient density gradients and ion-electron interaction is sufficiently weak (and this circumstance also depends on $m_e/m_i$ being very small, together with other conditions).

In the case of gyrokinetic transport form instabilities, the ramifications of eq(1) have some differences and similarities to this. As in the classical case, this results in very different behaviors to density gradients and temperature gradients. And like classical transport, the huge size difference in mass of electrons and ions is crucial to this differential behavior. But unlike the classical case, the fluctuations do not arise due to an unavoidable circumstance such as particle discreteness- but rather,. And the difference it the bahavior of temperature and density gradients is as follows. As has been shown in detail elsewhere, under particular circumstances, the density gradients together with the constraint eq(1) make it impossible for strong instabilities to even arise, even when there is great free energy from temperature gradients. This requires a large difference in the ratio $m_i/m_e$

We show this in two stages. Ultimately, we give the proof of eq(1) for exactly the gyrokinetic system examined in our simulations (which is widely used throughout the community): the ballooning limit. Within that limit, it is almost trivial, mathematically, to show that all the assumptions A-C can be dispensed with when deriving eq(1). But the very simplicity and triviality of the proof conceals the ultimate physical basis of the result. Since it is such an important constraint upon the dynamics of the fluctuations, here we give arguments here that reveal the physical origin, and the underlying kinship of these disparate cases. These arguments are an updating of those given by previously by Kotschenreuther [Ref [1], PhD thesis].

The well understood classical case is an excellent guide to the gyrokinetic case, once we adopt a general enough description that applies to both cases. This allows us to see that the same concepts can be used for both.

The theory of classical transport from Coulomb collisions in a strong magnetic field has been well understood for over half a century. It is accepted that the origin of eq(1) is "momentum conservation" in the collision process, defined in a specific way (Hinton and Hazeltine, [2]). Essentially the same definition of "momentum conservation" among the fluctuations giving the transport can be shown to hold for the gyrokinetic case. The classical Coulomb transport case has also been interpreted as arising from the general concept of elementary mechanics that the "sum of internal forces vanishes". This concept can be shown to be applicable to the gyro-kinetic case as well, once we suitably formulate the problem. For either binary interactions or collective gyrokinetic fluctuations, as long as they are small-scale: there is appropriate coarse-graining to scales greater than the scale of momentum transfers, and on these scales, momentum conservation holds in the same sense as it is used for classical transport.

A crucial element for the gyro-kinetic fluctuations is that they are a "quasi-steady state" of Maxwell's equations: the phase velocities are much less than the speed of light. This is also the same limit that applies to commonplace devices such as electric motors, capacitors, transformers, etc. Familiar concepts from electrostatics and magnetostatics thus apply, which allows an interpretation that the sum over internal forces vanishes, like the classical Coulomb collision case. (Other aspects of the gyrokinetic ordering are also helpful in this.)

For Classical transport, it is accepted that "momentum conservation" implies that the transport is "intrinsically ambipolar": that is, eq(1) always holds, no matter what the equilibrium parameters are (including, for arbitrary values of the radial electric field). One might expect that the fact that a concept as general as the sum over internal forces vanishes has a general applicability beyond binary collisions. And in fact, Coulomb collisions in a plasma already involve collective effects. As emphasized by Balescu [3], Lenard [4] and Gurnsey [5], the Coulomb collision operator in a plasma (the BGL operator) includes crucial collective interactions, where dielectric shielding clouds of the discrete particles play a key role in the "collisions". Thus, the BGL operator blurs the collision process into an interaction that includes both discrete particle elements (like the binary collision) and collective elements (dielectric shielding of the discrete particles). But, eq(1) holds nonetheless, because the Coulomb collision operator in a plasma still "conserves momentum", despite the collective interactions.

The gyrokinetic fluctuation case is the next step beyond the BGL operator, toward collective interactions. Collective (=dielectric) properties wholly determine the instabilities and fluctuations, which are no longer driven by particle discreteness. Particle discreteness plays only an indirect role by determining the collision operator in the gyrokinetic equation for the dielectric properties.

Nonetheless, the eq(1) still holds for gyrokinetic fluctuations because of the same essential elements: 1) a strong magnetic field to give small gyro-radius and 2) coarse-grained momentum conservation, for scales larger than the momentum exchange scale.

The essential element needed to generalize the type of argument employed in classical Coulomb collisional transport, we will see, is to merely interchange the order of the essential operations used in the classical arguments. Those essential operations for the classical case are, in their usual order, 1) coarse-graining. This is done in both velocity space and real space. This obtains the usual Coulomb "collision operator" for each species $C_s$, which is diffusive in velocity space but "local" (on the coarse-grained scale) in configuration space. In the next step 2) the momentum moment of this operator for each species, $C_s$ is taken, and it is summed over species s (i.e., one takes $\sum_s m_s \int d\vec{v}\, \vec{v}\, C$). This sum vanishes, and is interpreted as "momentum conservation" of the collisions. As we show below, by interchanging the steps 1) and 2), a major generalization of the concepts are possible.

By taking the velocity integral first, the coarse-graining step only needs to be applied in configuration space. The only difference between Coulomb collisions and gyrokinetic fluctuations is their scale in configuration space, and hence the scale of the coarse graining required. The same electromagnetic arguments apply to a very wide class of fluctuations. The interchange of the operations 1) and 2) above allows us to see that, insofar as the applicability of eq(1) is concerned, it is incidental that one set of fluctuations correspond to discrete particle effects (electrostatic Coulomb collisions) and the other corresponds to electromagnetic collective effects (gyrokinetic fluctuations).

A crucial question is, what is the relevant scale of the coarse graining? When the fluctuations are roughly isotropic, as in the classic Coulomb collisional case, the coarse graining scales must exceed the scale of the fluctuations (the Debye length.) But gyrokinetic fluctuations are very highly an-isotropic: they have an equilibrium scale in the direction parallel to the magnetic field $\vec{B}$, but far shorter scales perpendicular to $\vec{B}$. And here, an important point arises: insofar as momentum exchange is concerned, it turns out that the short scale determines the scale of perpendicular momentum exchange. It is unequivocally true that the long scale parallel dynamics determine exactly what the distribution function is, and hence, exactly what the momentum exchange between various species is. However, the spatial scale length of the relevant momentum exchange applicable to eq(1) is limited in all directions to the shortest scale. And therefore, the highly an-isotropic gyrokinetic case is equivalent to the isotropic Coulomb collisional case, insofar as eq(1) is concerned.

We demonstrate this latter point by two types of argument 1) general arguments based upon Maxwell's stress tensor and 2) a highly relevant limit of local plane waves. The latter is very pertinent to instabilities within the ballooning limit, which can be considered as a form of the WKB method (Wentzel, Kramers and Brillion) for waves with a short scale compared to equilibrium scales. The ballooning limit, like the WKB method, locally has short wavelength plane waves, and these waves change slowly in space. A plane wave is highly an-isotropic, just like gyrokinetic fluctuations: there is a short length in one direction and long (infinite) scale lengths in the others. Plane waves are a "text-book" case where the relevant calculations can be performed analytically. For a spherical volume average, coarse-grained momentum conservation holds when the radius is much larger than the wavelength. (After applying gyro-kinetic orderings of space and time scales.)

These arguments cement the interpretation of eq(1) as arising from essentially the same physical dynamics for classical Coulomb collisions and for gyrokinetic fluctuation transport (and likely other types of fluctuation transport as well). We now describe those steps in detail.

**II Review of classical transport**

Let us first briefly review the classical transport case, and then, see how the same considerations apply to gyrokinetic instabilities. We start with the momentum moment of the Vlasov equation. For simplicity, we ignore the inertia term as it applies to the slowly evolving equilibrium quantities, $\sim d\vec{v}/dt$. (This is formally justified in classical and neoclassical transport theory as an ordering, expanding in the limit where the gyro-radius is small compared to equilibrium scales. This describes evolution of the equilibrium on the transport time scale, where fast transients have been allowed to decay [2]. Similar orderings also apply when gyrokinetic fluctuation transport is present, since this transport scales as gyro-Bohm, like classical and neoclassical.)

The momentum moment of the Vlasov equation includes the momentum moment of the collision operator, $\vec{F}_s$, which gives the total frictional force from collisions:

$$\vec{\nabla} p + q_s n_s \vec{v}_s \times \vec{B} + \vec{F}_s + \vec{\nabla} \cdot \vec{\pi} + q_s n_s \vec{E} = 0 \qquad \text{Eq(2)}$$

The $\vec{v}_s$ here gives the transport flux (as well as diamagnetic flows, etc.). The classical transport flux results from the Coulomb collisional friction (HH1973):

$$n_s \vec{v}_{s\perp} = \vec{\Gamma}_{\perp s} = (1/q_s) \vec{F}_s \times \vec{B}/B^2 \qquad \text{Eq(3)}$$

For the classical transport case, eq(1) is valid because

$$0 = \sum \vec{F}_{\perp s} \qquad \text{Eq(4)}$$

Equation (4) is regarded as a statement of momentum conservation of the collisions. It holds because the collision events do not change the total momentum of the all the particles together, but merely redistributes it amongst those particles. Or stated in the language of elementary mechanics, the sum of all internal forces vanishes by Newtons third Law.

In this picture, Coulomb collisions are regarded as relatively local events. The corresponding collision operator is obtained by coarse graining in space (and time) over scales larger than the small-scale collision events. The Coulomb force is long-range, so these collisions involve particles separated by up to (roughly) a Debye length.

Obviously, some kind of spatial averaging is necessary to get momentum conservation: even for a simple binary Coulomb collision, one must use a volume that includes *both* colliding particles to have momentum conservation. See fig 1.

Coarse-graining over scales larger than the scale of momentum exchanges is crucial in order to get momentum conservation. The interactions do not create or destroy momentum, since they are internal forces of the system, but they transfer it to particles at a different location. When one coarse-grains over regions large compared to the exchange distance, these exchanges sum to zero, and the momentum is "conserved".

In addition, recall that coarse graining is always necessary in statistical mechanics to obtain the important behavior associated with entropy increase- such as obtaining transport fluxes from the collision processes or fluctuations. So coarse-graining over the scales larger than individual interactions is already part of the basic thermodynamic considerations.

To summarize, for classical transport, the well-known interpretation is that the flux constraint eq(1) results from momentum conservation of localized interactions, by coarse graining over scales that are larger than the scales of the momentum-exchanging events. Together with a strong magnetic field to make the gyro-radius small, eq(1) results. On the coarse-grained scale, eq(1) is locally true at every point along a flux surface, because, so is eq(4).

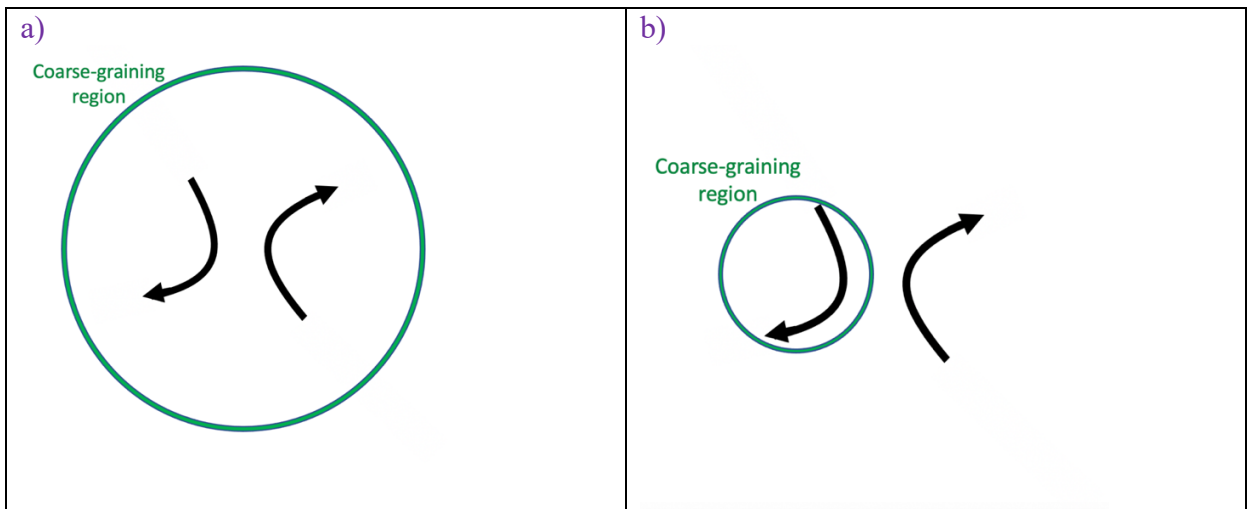

Fig 1. The concept of coarse graining a momentum exchange event over sufficient scales to obtain momentum conservation, displayed pictorially for a binary collision. The coarse-graining region in a) is large enough to give momentum conservation, since it includes the entire momentum exchange event. The region in b) is too small to give momentum conservation.

Other terms in eq(2) give the neoclassical transport from Coulomb collisions. However, gyrokinetic transport is most closely analogous to the classical case. This is at first surprising. It is well-known that gyro-kinetic fluctuations have a parallel scale of order the equilibrium. The same is true for the neoclassical contributions to eq(2). The neoclassical transport, eq(3) is true only after flux surface averaging- it is not local. One might suspect that similar considerations apply to the gyrokinetic case. Previous descriptions of the arguments here [Kotschenreuther, thesis 1982] employed a coarse-graining over an entire flux surface, analogously to neoclassical transport.

But eq(1) in fact holds locally for gyrokinetic fluctuations. This surprising result is because basic electrodynamic considerations (together with the gyro-kinetic ordering). The essential question is: over what kind of the volume does the average of the forces of the fluctuations vanish? This question can be addressed using Maxwell's stress tensor. We can apply this to anisotropic fluctuations, and obtain bounds using the Cauchy-Swartz inequality. The anisotropy of the gyrokinetic case offers no impediments to showing that the volume average of the relevant forces vanishes, even for a spherical volume. Another pertinent example is plane waves – which is what ballooning fluctuations are, locally. Calculations find that the volume average of perpendicular forces vanishes for spherical volumes.

On must distinguish between the scale lengths of the perpendicular momentum exchange and the scale lengths of other relevant dynamical processes. Essentially, the small wavelength of the gyrokinetic fluctuations in one direction is all that is needed to limit the length of perpendicular momentum exchanges-even though the dynamics of the fluctuations most definitely have a long parallel scale.

Let us now turn to gyrokinetic fluctuations.

**III Transport from gyrokinetic fluctuations**

By direct calculation of the nonlinear terms in the gyro-kinetic equation, for the standard theory in the ballooning limit, eq(1) is found to hold locally – just like classical transport. This is true whether or not the simplifications A-C apply. In the simple limit where A-C do apply, the fact that eq(1) holds locally is immediately obvious: because quasi-neutrality holds at every point (on a coarse grained scale), $\sum_s V_{E \times B} \, q_s \, \delta n_s = 0$. The reason why gyrokinetic transport is more analogous to classical transport, rather than neoclassical transport, goes back to the fundamental reason for eq(1): *the scale lengths over which perpendicular momentum is exchanged by the fluctuations is small compared to equilibrium scales.*

Before showing this, we must first write eq(2) in a form that explicitly includes fluctuations.

We split all quantities into mean parts (denoted by <>) and small fluctuating components (denoted by a $\delta$ ) that have zero average. The mean of eq(2) then contains additional terms that are quadratic in the fluctuations:

$$\vec{\nabla}<p> + q_s <\vec{\Gamma}_s> \times <\vec{B}> + <\vec{F}_s> + q_s <\delta n_s \delta\vec{E}> + q_s <\vec{\delta j}_s \times \delta\vec{B}>$$
$$+ \vec{\nabla}\cdot<\overleftrightarrow{\pi}> + q_s <n_s><\vec{E}> = 0 \qquad \text{Eq(5)}$$

In analogy to the classical collisional case, the flux due to the fluctuating quantities is

$$q_s \vec{\Gamma}_{\perp s} = \vec{F}_{fluctuation} \times <\vec{B}>/<B>^2 \qquad \text{Eq(6)}$$

Where $\vec{F}_{fluctuation}$ is the Lorentz force of the fluctuations, together with terms from the $\pi$ tensor.

$$\vec{F}_{s\,fluctuation} = q_s <\delta n_s \delta\vec{E}> + <\vec{\delta j}_s \times \delta\vec{B}> + \vec{\nabla}\cdot<\overleftrightarrow{\pi}_s>_{fluctuation} \qquad \text{Eq(7)}$$

Part of the $\pi$ tensor is included because there are components of $<\overleftrightarrow{\pi}>$ that are quadratic in the fluctuations ($\pi_{ij} \sim nm\delta v_i \delta v_j$). So we include this here, and we label it $<\overleftrightarrow{\pi}_s>_{fluctuation}$.

Eq(3) will hold when, for the components perpendicular to $<\vec{B}>$,

$$0 = \sum_s \delta \vec{F}_{s\,fluctuation} = <\delta\rho\,\delta\vec{E}> + <\vec{\delta j} \times \delta\vec{B}> + \sum_s \vec{\nabla}\cdot<\overleftrightarrow{\pi}_s>_{fluctuation} \qquad \text{Eq(8)}$$

Where $\delta\rho = \sum_s q_s \delta n_s$ and $\vec{\delta j} = \sum_s \vec{\delta j_s}$.

Eq(8) is, of course, is the exact analogue of the Coulomb collisional transport case.

Each term in eq(8) corresponds to one of the cases A, B and C outlines in section I.

A. The contribution of the first term in eq(S6) to the flux in eq(S5) has the appearance of the ExB drift flux, $(\vec{\delta E} \times \vec{B}/B^2)\delta n_s$. However, the $\vec{\nabla}\cdot<\overleftrightarrow{\pi}>_{fluctuation}$ contains gyro-viscous effects, and for fluctuations with $k_\perp \rho_i \sim 1$, its magnitude is comparable to the ExB term. Hence, when $k_\perp \rho_i \sim 1$, eq(4) does not obviously hold, even if the first term, alone, satisfies it.
B. The first term $<\delta\rho\,\delta\vec{E}>$ will not obviously vanish if the fluctuations are not quasi-neutral. This can arise, for example, for ETG modes, which can have scales of order the Debye length.
C. The second term, due to electromagnetic turbulent effects, also does not obviously vanish, even if the others do.

There are two common features of all the terms in eq(8). First, they are all self-forces of the plasma upon itself. Just as for the case of Coulomb collisions, we intuitively expect that such forces can redistribute momentum among particles, but cannot change the total momentum.

And second, all the terms can be written as the divergence of a tensor. In the case of the first two terms, this is the Maxwell stress tensor. This property is crucial once we coarse-grain the equations by taking a volume average.

For the electromagnetic terms:

$$< \delta\rho\, \delta\vec{E} > + < \delta\vec{j} \times \delta\vec{B} > = \vec{\nabla} \cdot \sigma^E_{ij} + \vec{\nabla} \cdot \sigma^B_{ij} + Poynting\ vector\ term \qquad Eq(9)$$

where $\delta\sigma^E_{ij}$ and $\delta\sigma^B_{ij}$ are the electric and magnetic part of the Maxwell stress tensor

$$< \delta\sigma^E_{ij} > = \frac{1}{4\pi} < \delta E_i \delta E_j - \frac{1}{2} I_{ij} \delta E^2 > \qquad Eq(10)$$

$$< \delta\sigma^B_{ij} > = \frac{1}{4\pi} < \delta B_i \delta B_j - \frac{1}{2} I_{ij} \delta B^2 > \qquad Eq(11)$$

(The fluctuating Poynting vector momentum is small, for both gyrokinetic fluctuations and the usual Coulomb collision case. We consider the former momentarily.)

Since all the terms are a divergence, the volume integral becomes a surface integral. Because the ratio of surface area S to volume V approaches zero as the scale size of the volume increases, a volume average of these terms is $\sim S/V \to 0$ as the coarse graining volume increases. How big does the volume need to be? We estimate this, in a moment, for the gyrokinetic case.

But first, we point out the relationship of eq(8) to the classical Coulomb collisional case.

**Comparison to the classical Coulomb collisional transport: interchanging the order of coarse graining and velocity integration to obtain a much more general argument**

But first, since the Coulomb collisional case is such an instructive analogy to the gyrokinetic case, we consider how it, too, can be obtained from the fluctuation term in eq(8). The Coulomb collision operator can be obtained directly from the Vlasov equation as indicated by Klimontovich [Klimontovich reference]. He chose the distribution function f to be a sum of delta functions at the position and velocity of each discrete particle, $f(\vec{x}, \vec{v}, t) = \sum_i \delta(\vec{x} - \vec{x}_i(t)) \sum_i \delta(\vec{v} - \vec{v}_i(t))$. Since the particles (labeled by i) obey the Lorentz force, this distribution function obeys the Vlasov equation, and its moment is eq (8). Of course, the Vlasov equation for this distribution function does not have a collision operator, since it describes each individual particle- it has not yet been coarse grained. But this distribution function can be used to *derive* the usual Coulomb operator after standard coarse graining and statistical methodologies are adopted. This results in the usual Coulomb collision operator that is diffusive in velocity space and local in coordinate space. It is standard nomenclature to say that the Coulomb collision operator results from the coarse-grained effects of fluctuations arising from "particle discreteness".

What one should take away from this example is that eq(4) can be obtained, for Classical transport, from the Vlasov equation (Klimontovich version). The *traditional* order of operations is to 1) first coarse grain to obtain the collision operator $C_s(f)$, and then 2) take the velocity moment times species mass and then sum over species, $\sum_s m_s \int dv\, \vec{v}\, C_s(f)$. This leads to the constraint eq(4) and hence eq(1).

*It is highly instructive to consider the result of interchanging operations 1) and 2) above.* In other words, we could *first* take the velocity moment of the Klimontovich distribution function. Then, take the integral over v, multiply by the mass, and sum over species. Then, *secondly*, we *start* to perform a coarse graining by setting up the *definitions* of coarse grained and fluctuations. The first step of this is to define the coarse grained distribution function, which is smooth, as are it's moments. Then, there are fluctuations around it, which, for classical transport, due to the discrete nature of particles. At this point, eq(5) holds, but with no explicit collision operator C or terms $<\vec{F_s}>$. The effect of particle discreteness is fully contained in the terms $q_s <\delta n_s\, \delta \vec{E}>$. In other words, for the Klimontovich distribution, the $q_s <\delta n_s\, \delta \vec{E}>$ is equal to $<\vec{F_s}>$, *after* a more elaborate next step to obtain a the usual Fokker-Plank operator from the coarse-grained effects of the fluctuations.

*The crucial point is that, if one's sole purpose is to proving eq(1) holds, it is not necessary to actually apply coarse graining in velocity space to explicitly obtain a collision operator C. One can apply coarse graining directly to eq(8) in coordinate space. And this same procedure can be applied for any fluctuations- such as gyrokinetic ones.*

For the classical transport case, we expect that this transposed order of procedures must give the same result as the for the standard order of coarse graining, followed by taking $\sum_s m_s \int d\vec{v}$. In other words, eq(7) must coarse-grain to zero when the fluctuations arise from particle discreteness. This is indeed the case.

*However, when we start from eq(7), the fundamental relationships that one uses for the classic Coulomb collisional case to show that eq(8) follows are, in fact, the same as can be used for the case of gyrokinetic fluctuations (or other types of fluctuations too). The only thing that changes is the scale of the fluctuations.*

**The coarse-graining arguments in configuration space**

We now describe these arguments in detail, in two stages:

1) We give them in their most general form, using Maxwell's stress tensor.
2) We describe an illustrative and highly relevant example: the case of plane waves.

Plane waves are a good approximation to gyrokinetic modes in the ballooning limit. The ballooning limit is WKB-like: the fluctuation locally has a large $\vec{k}$ vector which changes on a scale much longer than $1/k$. So, fluctuations are locally plane waves. So this example is quite representative of the actual gyrokinetic case where we use the constraint to explain the stability properties of the modes.

This case is a "textbook" example in classical electrodynamics, where everything can be readily calculated. *In particular,* it has a short scale only in one direction, and an infinite correlation length in others- the character of gyrokinetic fluctuations. For a spherical volume, the relevant volume averages vanish when $kR \gg 1$.

Other important characteristics can readily be shown by application of the gyrokinetic equation:
3) In the gyrokinetic ordering, the fluctuations are in the electrostatic and magnetostatic limit.
4) The Poynting vector momentum is small
5) So, the system can be described in terms of forces which are like elementary Newtonian mechanics: results can be interpreted as the sum of internal forces vanishing.
6) The same concepts can be used for electromagnetic gyrokinetic fluctuations as for classical case of transport from electrostatic Coulomb collisions

These cases allow us to demonstrate that if fluctuations have a short scale in any direction, the relevant electromagnetic momentum exchange is limited that similar scale. The momentum exchanges from $\sum_s \vec{\nabla} \cdot <\overleftrightarrow{\pi}_s>_{fluctuation}$, relative to the $<\delta\rho\, \delta\vec{E}>$ term, are limited in scale by particle gyro-motion.

This in no way contradicts the fact that the gyro-kinetic distribution function is determined by long-scale dynamics along a field line. *The constraint upon the scale of momentum exchange arises no matter what the distribution function is, or how it is determined.* On must distinguish between the scale lengths of the perpendicular momentum exchange and the scale lengths of other dynamical processes.

The fact that the actual perturbed distribution function determined by relatively long-range dynamics along $\vec{B}$ is immaterial: the limitation on the scale length of momentum exchange holds regardless, as shown in the following calculations. And so, eq(1) holds locally (in the coarse grained sense), since it is an consequence of coarse grained momentum conservation, averaged over scales larger than the momentum exchange.

Pictorially, we can represent this arguments as an analogy to fig 1. These figures are helpful in orienting one to the geometry.

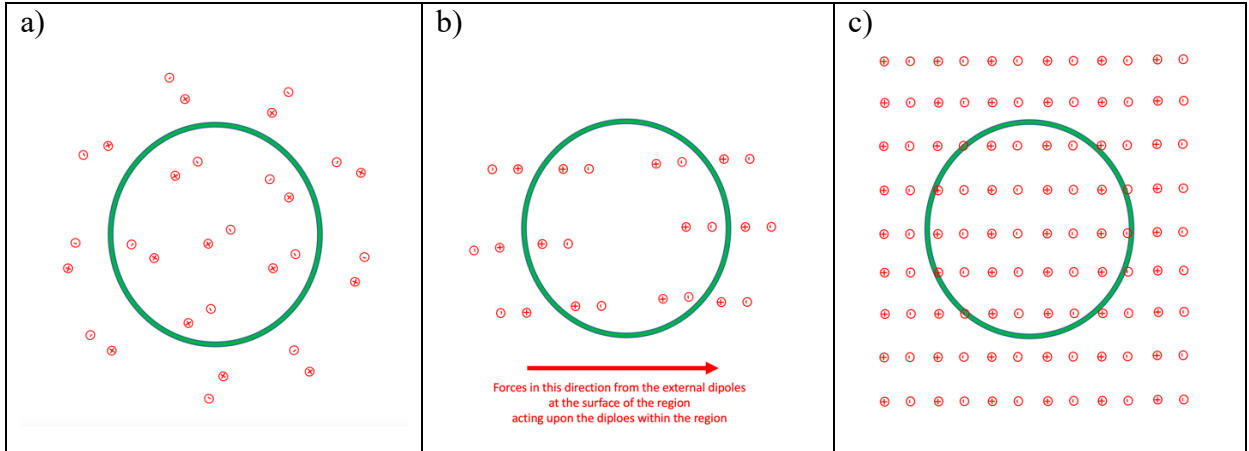

Fig 2. Forces from fluctuation charges $\delta\rho\, \delta\vec{E}$ represented as dipoles, as arise in gyrokinetic fluctuations. Since the fluctuations have zero average, fluctuation charges could be conceptualized as dipoles, with a distance between charges ~ 1/k. The sum of forces from charges inside the region on themselves always vanishes. *A net force on the charges inside the region can only arise from charges outside.* a) **For gyrokinetic turbulence, the net forces** from randomly oriented dipoles outside is sometimes in one direction and sometimes in the opposite direction. The sum of randomly cancelling forces is small when R >> 1/k. b) To get the maximum force possible on the region, the dipoles would all line up to always give the same force direction. Even in this case, *since the effective range of forces of dipoles outside is of order the dipole separation  ~ 1/k, Net force ~surface area × range ~ S/k*. Hence the volume averaged $force \sim S/kV$ vanishes as the ratio of surface area to volume. For a circular region with radius R, for large R, this vanishes as ~ $1/kR$. We obtain this same scaling for the maximum volume averaged force by analytic arguments using Maxwell's stress tensor, in the text. c) The case of a plane wave. Although this is highly ordered, it is still true that the external forces from the dipoles outside on the charges inside are sometimes in one direction and sometimes in the opposite direction. Calculation in the text shows that the volume average force ~ $1/(kR)^2$

Note that the pictures in Fig 2 apply if the charges are highly elongated charge rods in the direction out of this plane. This is the case of an-isotropic gyrokinetic turbulence, with a parallel scale much greater than the perpendicular scale. In this case, for any 2D plane perpendicular to the rods, the average perpendicular forces still vanish as the radius of the region exceeds the typical charge separation distance, which is ~ $1/k_\perp$. Since this holds for any perpendicular plane, it is also true of a cylindrical volume of any height. That height could be of order the radius- that is, the volume average force still vanishes even if the volume is not highly elongated like the charge rods. And the same picture holds if this electrostatic picture is replaced by a magnetic one where the charge rods are replaced by current carrying rods, with currents in and out of the page. This corresponds to magnetic forces from $\delta\vec{j} \times \delta\vec{B}$ interactions in the gyro-kinetic case.

**General arguments and applications of the gyrokinetic ordering, using Maxwell's stress tensor.**

We define a local geometry where $<\vec{B}>$ is in the z direction, x is the direction across flux surfaces, and y within the flux surface and normal to $<\vec{B}>$.

Then the relevant component of eq(8) is

$$<\delta\rho\, \delta\vec{E}> \cdot \hat{y} = \vec{\nabla} \cdot <\delta\sigma^E_{ij}> \cdot \hat{y} \qquad \text{Eq(12)}$$

So the volume $\{\}_{vol}$ average of eq(9) becomes a surface integral

$$\{<\delta\rho\, \delta\vec{E}> \cdot \hat{y}\}_{vol} = \frac{\int d\vec{x} <\delta\rho\, \delta\vec{E}> \cdot \hat{y}}{V} = \frac{\int d\vec{S} \cdot <\delta\sigma^E_{ij}> \cdot \hat{y}}{V} \qquad \text{Eq(11)}$$

We can now use several applications of the Cauchy-Swartz inequality $|\vec{a} \cdot \vec{b}| \leq (\vec{a} \cdot \vec{a})^{1/2}(\vec{b} \cdot \vec{b})^{1/2}$, and the algebraic inequality $|x + y| \leq |x| + |y|$, to arrive at

$$\{<\delta\rho\, \delta\vec{E}> \cdot \hat{y}\}_{vol} \leq \frac{3}{8\pi} \frac{\int dS\, \delta E^2}{V} \qquad \text{Eq(12)}$$

Let us compare this volume average to the "nominal" size of $<\delta\rho\, \delta\vec{E}>$. Since $\vec{\nabla} \cdot \delta\vec{E} = 4\pi\delta\rho$, if we have a typical wavenumber k, then $<\delta\rho\, \delta\vec{E}>_{typical} \sim \frac{k}{4\pi} \delta E^2$. Then the ratio of the volume averaged value to the nominal, or typical, value is

$$\frac{\{<\delta\rho\, \delta\vec{E}> \cdot \hat{y}\}_{vol}}{<\delta\rho\, \delta\vec{E}>_{typical}} \leq \frac{3}{2} \frac{\int dS\, \delta E^2}{kV\, \delta E^2} \qquad \text{Eq(13)}$$

For the turbulence of interest, $\delta E^2$ is fairly uniform. It varies on a scale much longer than $1/k$. This is true nonlinearly, where the turbulence is driven locally by the equilibrium gradients. And, linear modes in the ballooning limit are also comparatively longer scale than $1/k$ (which may be an intermediate scale between the gyroradius and the equilibrium). Then the $\delta E^2$ in the numerator roughly cancels the one in the denominator, and we have:

$$\frac{\{<\delta\rho\, \delta\vec{E}> \cdot \hat{y}\}_{Vol\, avg}}{<\delta\rho\, \delta\vec{E}>_{nominal}} \lesssim \frac{3}{2} \frac{S}{kV} \qquad \text{Eq(14)}$$

If the volume has a characteristic dimension of R, as in a sphere, coarse graining reduces the force from its typical value by $\sim (Rk)^{-1}$. In other words, as the coarse graining dimension exceeds the characteristic fluctuation dimension, the coarse-grained flux vanishes.

*The fact that k may be large only in the perpendicular direction does not change this result.*

How does this compare to the nominal size of the flux in the gyrokinetic ordering? We use the usual gyrokinetic orderings $\delta n/n \sim e\delta\phi/T$, $\delta E \sim k_\perp \delta\phi$, $k_\perp \rho_s \sim 1$, $k_\perp^2 \delta\phi \sim \delta\rho$, $\delta v_{E\times B} \sim (c/B) k_\perp \delta\phi$. We compare the nominal charge flux in gyro-kinietcs $\sim e\delta n\, \delta v_{E\times B}$ to the results for the volume averaged charge flux implied by the above. We find that

$$\frac{\{\hat{y}\cdot\Sigma_s q_s<\delta n_s\,\delta\vec{E}>\}_{Vol\,avg}}{q<\delta n\,\delta\vec{E}>_{nominal}} \sim k^2\lambda_{Debye}^2 \frac{S}{kV} \qquad Eq(15)$$

As described in section I, when $k\lambda_{Debye} \ll 1$, coarse graining in not needed to make this small. Quasi-neutrality does this. However, some modes can have $k\lambda_{Debye} \sim 1$, such as ETG modes. The validity of eq(1) then requires coarse graining to make $S/kV$ small.

Now, consider finite gyro-radius terms, which rise from $<\overleftrightarrow{\pi}_s>_{fluctuation}$. We can estimate this as $\pi_{ij} \sim nm\delta v_{E\times B\,i}\delta v_{E\times B\,j}$, and use the standard gyrokinetic orderings above. We can also estimate this from the nonlinear part of the gyro-viscous tensor, which gives the same result for the magnitude. We compare this to the size of the nominal gyrokinetic flux. We obtain

$$\frac{\{\hat{y}\cdot\Sigma_s q_s\vec{\nabla}\cdot<\overleftrightarrow{\pi}_s>_{fluctuation}\}_{Vol\,avg}}{q<\delta n\,\delta\vec{E}>_{nominal}} \sim \frac{\rho_s S}{V} k_\perp\rho_s \qquad Eq(16)$$

When $k_\perp\rho_s \ll 1$, this is small even without coarse-graining. But when $k_\perp\rho_s \sim 1$, coarse-graining is required in order for eq(1) to hold.

The magnetic term behaves roughly analogously to the electrostatic one. In the usual maximal ordering, the magnetic terms in the gyrokinetic equation have the same size as the electrostatic ones. For example, $q_s\delta n\,\delta\vec{E} \sim \delta\vec{J_s}\times\delta\vec{B}$. By using the magnetic part of the Maxwell stress tensor, very similar arguments can be used as for the electric terms. We have

$$\frac{\{<\delta\vec{j}\times\delta\vec{B}>\cdot\hat{y}\}_{Vol\,avg}}{<\delta\vec{j}\times\delta\vec{B}>_{nominal}} \lesssim \frac{3}{2}\frac{S}{kV} \qquad Eq(17)$$

This only vanishes because of coarse-graining.

In conclusion, these results correspond to the three cases A-C in section I. In all cases, the volume average is needed in order to arrive at eq(1), except when A-C hold.

We now turn to some other consequences of the gyrokinetic ordering.

<u>The smallness of the Poynting Vector</u>

The Poynting vector term for the electromagnetic momentum was neglected in eq(9). Within gyro-kinetics, the nominal, maximal ordering is for the electromagnetic effects to be of order the electrostatic effects. For example, this implies $\delta\phi \sim (v_{th\,e}/c) A_\parallel$. Other nominal orders are $\omega \sim \frac{v_{th\,i}}{L}$ for ion-scale modes ($k_\perp\rho_i \sim 1$) and $\omega \sim \frac{v_{th\,e}}{L}$ electron scale modes (ETG) ($k_\perp\rho_e \sim 1$). These imply that the Poynting vector term, compared to the electric term in eq(9), is $\sim (m_e/m_i)^{\frac{1}{2}} (\rho_i/L)$ for ion scale modes and $\sim (\rho_e/L)$ for electron scale modes. These are both small in the gyrokinetic ordering. Hence, the Poynting vector is neglected in eq(9).

Quasi-static limit

Gyro-kinetic fluctuations are in the quasi-static limit of Maxwell's equations. One can easily interpret the vanishing of the volume averaged forces eq(9) as the sum over internal forces vanishing.

This is the basic conceptual parallel between the Coulomb collisional case and the gyrokinetic case.

The quasi-static lit also applies to everyday devices such as transformers, capacitors, electric motors, etc., and we can apply some of those basic results to the present problem.

Gyro-kinetics are in the quasi-static limit since the phase velocities of the fluctuations are much smaller than the speed of light, that is, $\omega/kc$ is small. Within the gyrokinetic ordering this is $\omega/kc \sim \rho/L$ which is the small ordering parameter.

In a quasi-static limit, the potential $\delta\phi$ and $\delta\vec{A}$ are determined by Poison's equation and Ampere's Law, and $\delta\vec{B}$ can be found from the Biot-Savart Law.

$$\delta\phi(\vec{r}) = \int d\vec{r'}\ \delta\rho(\vec{r'})/|\vec{r}-\vec{r'}| \qquad \text{Eq(18)}$$

And

$$\delta\vec{B}(\vec{r}) = \int d\vec{r'}\ (\vec{r}-\vec{r'}) \times \delta\vec{j}(\vec{r'})/|\vec{r}-\vec{r'}|^3 \qquad \text{Eq(19)}$$

First, note that in the gyro-kinetic ordering, the perpendicular $\vec{E}$ is from the electrostatic potential, and the inductive $\vec{E}$ is only for the parallel direction. (Within the language of MHD, shear Alfven waves are part of gyro-kinetics, but compressional waves are not since they are too fast for the ordering.) Hence, the electric fluctuation term of eq(8-9)is

$$<\delta\rho\ \delta\vec{E}> = <\delta\rho\ \vec{\nabla}\delta\phi>$$

We use eq(18) to compute the forces only from the charges inside a coarse-graining volume V

$$\int_V d\vec{r}\ \delta\rho(\vec{r})\vec{\nabla}\delta\phi(\vec{r}) = \int_V d\vec{r} \int_V d\vec{r'}\ \frac{\delta\rho(\vec{r})(\vec{r}-\vec{r'})\delta\rho(\vec{r'})}{|\vec{r}-\vec{r'}|^3} = 0 \quad \text{Eq(20)}$$

This force vanishes because the integrand is antisymmetric under the interchange $\vec{r} \leftrightarrow \vec{r'}$. *Or physically, the force of the charges at $\vec{r}$ on those at $r'$ are equal but opposite to the forces at $r'$ from those at $\vec{r}$. This is just Newton's third law.*

This is just the extension of the concept from binary collisions to the multi-particle case of collective instabilities where the charges are regarded as distributed in a continuum. The sum of internal forces vanishes.

The Biot-Savart Law gives exactly the same result for magnetic forces of currents within the volume V on themselves.

*Because of there are no integrated force inside V from charges and currents inside it, such a force can only arise from the charges and currents outside of V.* If the scale of the fluctuations is $\sim k$, then the fields from outside V only penetrate into V a distance $\sim 1/k$. Hence, the volume average force on the volume must vanish at least as quickly as the surface to volume ratio

$$Volume\ average\ force \sim \frac{S}{kV} \qquad \text{Eq(21)}$$

This, of course, is exactly the result of the bounding estimates from the Maxwell stress tensor.

If the forces from the boundary are no all in the same direction, but have contributions with different signs, the forces can cancel, so that the forces can be less than the bounding estimate. This is exactly the case in Fig 2a and Fig 2c. We now turn to the plane wave case.

**The example of plane waves- pertinent to the ballooning limit**

We now consider the case of plane waves. First, we describe why this is especially relevant to gyrokinetic instabilities in the ballooning limit

Recall the ballooning limit ansatz is that perturbed mode quantities $\delta f$ behave as

$$\delta f = \delta f(\psi, \theta, \zeta) e^{in(\zeta - q(\psi)\theta)} \qquad \text{Eq(22)}$$

For large n. Thus, ignoring lower order terms,

$$\vec{\nabla} \delta f = i\vec{k}\ \delta f \qquad \text{Eq(23)}$$

where $\vec{k} = n\vec{\nabla}(\zeta - q(\psi)\theta)$. The wavenumber $\vec{k}$ is large and varies on the equilibrium space scale. To lowest order, $\delta f$ is locally a plane wave with a short wavelength.

A plane wave has a short wavelength in only one direction, and an infinity large space scale in the others. So, it has qualitatively the strong an-isotropy of gyrokinetic fluctuations.

For a plane wave, it is trivially possible to solve for the fields from a given charge and current. For the electrostatic force $\delta\rho\ \delta\vec{E} \cdot \hat{y}$, $4\pi\delta\rho = i\vec{k} \cdot \vec{E}$. Let us compute the volume average of this over a spherical region of radius R.

A sphere has no preferred direction, so the results depend only upon the magnitude of $\vec{k}$. The nominal value of the force is $\sim \delta E_y\ \vec{k} \cdot \delta\vec{E}/4\pi$. We use the rules of complex manipulations for real physical quantities, and suppose $\delta E$ has an particular phase $e^{i\theta}$. The ratio of the nominal value to the volume average, for $kR \gg 1$, is computed to be

$$\frac{\{\delta\rho\ \delta E_y\}_{Vol\ avg}}{\delta\rho\ \delta E_{y\ nominal}} \rightarrow \left(\frac{3}{2}\right)\frac{\sin(2\theta)\cos(kR)}{(kR)^2} \qquad Eq(24)$$

The magnitude of this expression decreases as $\sim(kR)^{-2}$, even faster than the bounding estimate computed from the Maxwell stress tensor, which decreased as $\sim 1/(kR)$. Recall that the bounding estimate used the Cauchy-Swartz inequality and other bounding inequalities, so this results is consistent with it being an upper bound.

The magnetic case similarly obeys (using gyrokinetic orderings):

$$\frac{\{\hat{y}\cdot\delta j\times \delta B\}_{Vol\ avg}}{\hat{y}\cdot\delta j\times \delta B_{nominal}} \rightarrow \left(\frac{3}{2}\right)\frac{\sin(2\theta)\cos(kR)}{(kR)^2} \qquad Eq(25)$$

We estimate that the $\vec{\nabla}\cdot<\overleftrightarrow{\pi}_s>_{fluctuation}$ term also decreases $\sim(kR)^{-2}$.

**Discussion of the results**

So in the case of Coulomb collisions, there is no contradiction between the proof we have just given, based upon Maxwell's stress tensor, and the usual interpretation given for the vanishing of $0 = \sum \vec{F}_{\perp s}$ for the collision operator due to momentum conservation in the collision process. Both can be interpreted as being a consequence of the elementary fact that the sum over internal forces vanishes. As long as one can coarse grain over a volume large compared to the interaction distance, the only net force upon the volume are from interactions with particles outside the volume. And this must vanish by the ratio of the surface area to the volume, when the dimensions of the volume exceed $\sim\lambda$. For example, the volume can be a sphere of radius R, and $R \gg \lambda$.

In the case of Coulomb collision, the relevant scale of the fluctuations is the Debye length. This is the cut-off to the interaction distance in the usual derivations of the Coulomb collision operator.

But the arguments above by coarse-graining in configuration space are extremely general, and they apply to gyrokinetic fluctuations just as for the Coulomb collisional case. *It is just that the scale of coarse graining must be commensurate with the fluctuation scale. So it applies, in the case under consideration, to gyrokinetic fluctuations. And furhter, the an-isotropy of the gyrokinetic fluctuations does not affect the applicability of the arguments, nor, does the fact that the gyrokinetic fluctuations have magnetic forces as well as electric ones.*

As for the $\pi$ term, the gyromotion spreads the particle over a gyroradius, so one can expect the viscous forces inside and outside the volume to mix over that distance. So it is not surprising that the coarse graining must be over distances large compared to gyroradius in order for eq(3) to hold.

So, to summarize: when one applies coarse graining, the self-forces in eq(7-8) vanish as the volume exceeds the scale of the fluctuations and the gyroradius. This is conceptually the same argument as is applied in the case of classical Coulomb collisional transport, and for the same reasons. It is just that the scale needed for appropriate coarse graining is different, as one would expect from the different scales of the interactions.

**The proof of the constraint eq(3) directly from the gyro-kinetc equation.**

With these concepts in mind, we now consider the gyrokinetic equation in the ballooning limit. There are derived expression for the fluxes in this gyrokinetic limit. Since the gyrokinetic equation is just a specific order of the Vlasov equation, all the same results must apply. And they do. Though without the conceptual arguments just given, it would not be obvious that the gyrokinetic case is, fundamentally, the same as the case of classical transport.

The eq(1) is a straightforward consequence of the gyrokinetic equation. In all the cases described below, eq(1) holds locally, just as in classical transport. Just as in the arguments above from the Vlasov equation and Maxwell's equations, one does not need to average over an elongated volume, or a flux surface, to arrive at eq(1).

The gyrokinetic equation describes the distribution of gyro-centers. The ballooning limit is a type of WKB-like approximation. It is the most widely used equation to describe gyrokinetic turbulence. At each position along a field line, each wave has a particular $\vec{k}_\perp$ as a specified function of position. Nonlinear quantities are computed by using the usual rules for summing Fourier terms.

Though this leads to some algebraic complexity, but makes the calculations completely straightforward.

Let us consider electrostatic fluctuations first. The gyro-centers move according to the *gyro-orbit averaged ExB drift*, which is well known to become

$$\vec{v}_{gyro} = (\vec{\delta E} \times \vec{B}/B^2) J_0 ( k_\perp \rho_s ) \qquad \text{Eq(26)}$$

Where $J_0$ is the Bessel function of order zero.

For $k_\perp \rho_s \sim 1$, these velocities vary considerably for different species and energy. Because of this, it is not at all obvious, a-priori, that the total charge flux will vanish. This corresponds to case A above.

The transport flux of gyro-centers is the product of this velocity with the distribution function of gyrocenters. It is conventionally to use the non-adiabatic part of the distribution, $\delta h$. The flux of gyrocenters is

$$Gyrocenter\ flux = \vec{v}_{gyro}\ \delta h$$

To obtain the particle flux we integrate over velocity. The transport flux of gyrocenters that applies to the equilibrium transport is the $\vec{k} \rightarrow 0$ part of the nonlinear flux.

Crucially, note that this procedure, which is used throughout the community for the calculations of transport fluxes in the equilibrium, is a form of coarse-graining. It is equivalent to the volume average above for the limit of large volume dimension.

The density perturbation in real space is related to the density of gyrocenters by

$$\delta n_{k_\perp} = \frac{e\phi_{k_\perp}}{T} + \int dv\ \delta h_{k_\perp} \qquad \text{Eq(27)}$$

This is the density used in Poisson's equation in the gyrokinetic system. Finally, we use $\overrightarrow{\delta E}_{k_\perp} = i\overrightarrow{k_\perp}\ \delta\phi_{k_\perp}$. Consider first the limit of this when the Debye length is small compared to the fluctuation scale, that is, when quasi-neutrality applies.

Using the usual rules for products of Fourier quantities, it is straightforward to show that eq(1), the vanishing of the charge flux, holds. The $J_0(\ k_\perp \rho_s)$ that appears in the eq(26) is exactly what is needed in eq(27) so that quasi-neutrality implies eq(1).

Within the gyrokinetic framework, it is a rather remarkable algebraic coincidence that the $J_0$ factor in the orbit average of the ExB drift, and for the quasi-neutrality condition, conspire to lead to eq(1).

In view of the arguments in the section above, the interpretation is that, since gyrokinetics is a systematic and consistent expansion of the Vlasov equation, eq(1) must hold. So, some such "remarkable coincidence" was foreordained.

Now, let us dispense with the quasi-neutrality and consider fluctuations of the scale of the Debye length, as well as finite $k_\perp \rho_s$: as occurs for ETG turbulence, for example.

Then Poison's equation becomes

$$-k_\perp^2 \phi_{k_\perp} = \sum q_s \delta n_{k_\perp} \qquad \text{Eq(28)}$$

Using the usual rules for the product of Fourier quantities, and using the usual expression for the nonlinear flux in gyrokinetics, the sum of the particle fluxes times charge in the radial (x) direction (where y is the direction perpendicular to the magnetic field and to x):

$$\sum q_s \, \vec{\Gamma}_{xs} =$$

$$Re(\left(\tfrac{c}{B}\right) ik_y \, \phi_{\vec{k}} \sum q_s \delta n^*_{\vec{k}}) =$$
$$\left(\tfrac{c}{B}\right) Re(\, ik_y \, \phi_{\vec{k}} \, (-k_\perp^2 \phi^*_{\vec{k}} \,)) = \left(-\tfrac{c}{B}\right) Re(\, ik_y \, k_\perp^2 (\phi_{\vec{k}} \, \phi^*_{\vec{k}}) \,) \qquad \text{Eq(29)}$$

Of course, $(\phi_{\vec{k}} \, \phi^*_{\vec{k}})$ is real, and $k_\perp^2$ are purely real. Therefore, this vanishes by trivial algebra.

So eq(1) still holds even without quasi-neutrality, because for "some reason", when the electrostatic parts of Maxwell's equations are applied, the flux is the real part of a quantity that is purely imaginary. There is no obvious physical interpretation of this algebraic result within the the gyrokinetics formalism, it simply arises, almost trivially, by "turning the crank" within the formalism.

Let us compare all the features of this result with the quite different arguments in the section above.

Note that eq(29) is for the $\vec{k} = 0$ Fourier component of the charge flux, so that the nonlinear product is for the beat together of two terms with the same $\vec{k}$ (or, what is equivalent for a real physical quantity such as $\delta\phi$ or $\delta n$, the beat of $\vec{k}$ and $-\vec{k}$). It does not hold for the $\vec{k} \neq 0$ part of the charge flux. In fact, zonal flows are generated precisely because the nonlinear $\vec{k} \neq 0$ charge flux does not vanish.

This is mathematically equivalent to saying that it is only the volume average flux that vanishes, in the limit of large volume. Or using the usual terms of thermodynamics, it only vanishes in the coarse-grained sense.

Notice also that eq(29) holds for every position along a field line. In other words, the coarse-grained charge flux vanishes at every point. Again, this is like classical transport, where eq(1) holds locally, and does not require a flux surface average to hold.

To summarize, since gyro-kinetics is a systematic and consistent expansion of the Vlasov equation, then eq(1) holds within that limit. The physical interpretation in terms of the Vlasov equation, that was described in the section above, holds. The Vlasov equation version is much clearer physically; for example, it can easily make connections to classical Coulomb collisional transport. Also, the Lorentz force terms and the application of Maxwells equations is simpler, since these so not have to be filtered through the gyrokinetic ordering. The result, eq(1), does indeed hold in that limit, as can be seen by evaluating the terms.

So eq(1) can be said to arise for the gyrokinetic equation from coarse-grained momentum conservation, in exactly the same sense as it arises for classical Coulomb collisions.

The electromagnetic terms in the gyrokinetic flux follow this same pattern. For example, magnetic stochasticity leads to fluxes across field lines. Within the gyrokinetic ordering, this arises from terms due to the parallel component of $\vec{A}$, or $A_\parallel$. Using the standard expressions for nonlinear flux within gyrokinietcs the coefficient of $A_\parallel$ in Ampere's law is $-k_\perp^2$, which is real, eq(1) holds. While this is easy to establish algebraically, there is no obvious physical interpretation. Once again, this is provided by the discussion following the Vlasov equation results.